\documentclass[
 reprint,
superscriptaddress,
 amsmath,amssymb,
 aps,
 prx,
]{revtex4-1}

\usepackage{graphicx}
\usepackage{dcolumn}
\usepackage{bm}

\begin{document}

\title{Dynamic Nuclear Polarization by optical Stark effect in periodically-pumped gallium arsenide}

\author{Michael J. Dominguez}
 \affiliation{Applied Physics Program, University of Michigan, Ann Arbor, Michigan 48109, USA}
 
\author{Joseph R. Iafrate}
\affiliation{Applied Physics Program, University of Michigan, Ann Arbor, Michigan 48109, USA}

\author{Vanessa Sih}
 \email{vsih@umich.edu}
 \affiliation{Department of Physics, University of Michigan, Ann Arbor, Michigan 48109, USA}


\begin{abstract}
Optical pump-probe time- and magnetic-field-resolved Kerr rotation measurements provide a window into the mechanisms that generate dynamic nuclear polarization in bulk gallium arsenide. Previously, we have reported an unexpected dependence of the direction of the nuclear polarization on the sweep direction of the applied external magnetic field. In this paper, we present numerical calculations based on a model for this nuclear polarization due to the optical orientation and optical Stark effect produced by a train of ultrafast optical pulses. We demonstrate the correspondence of the model to our experimental measurements for different laser wavelengths and magnetic field sweep rates. Finally, we show that the model reproduces the sweep direction dependence and provides an explanation for this behavior.
\end{abstract}

\maketitle

\section{\label{sec:intro}Introduction}

The study of electron spin polarization in solids has been motivated by potential applications in quantum and classical spin-based computation and has revealed intriguing interactions with the nuclear spin system \cite{Urbaszek, Chekhovich}. Nuclear spin polarization manifests as an effective magnetic field, the Overhauser field, which affects the Larmor precession of the electron spin polarization. Measuring the Larmor precession frequency of electron spin polarization over time provides insight into the buildup of dynamic nuclear spin polarization (DNP).

All-optical studies have observed DNP in bulk semiconductors \cite{Lampel, Kikkawa2000, Zhukov2014, Heisterkamp}, quantum wells \cite{Salis, Zhukov2014}, and quantum dots \cite{Gammon, Markmann}. Pumping of electron spin polarization with a train of optical pulses allows for resonant spin amplification (RSA), in which electron spin polarization due to successive optical pulses builds constructively \cite{Kikkawa1998}. This enhancement in electron spin polarization is periodic with respect to the transverse magnetic field amplitude. Measurements in fluorine-doped ZnSe \cite{Zhukov} and GaAs \cite{Macmahon} have both revealed a DNP that is periodic in magnetic field. The former was attributed to electron spins optically rotated onto the magnetic field axis via the optical Stark effect (OSE) \cite{Kopteva}, while the cause of the latter was not conclusively identified.

Furthermore, the latter DNP was characterized by a dependence on the sweep direction of the external magnetic field. By combining time-resolved and magnetic-field-resolved scans, we extracted the Overhauser field experienced by the electron spin system. The sign of the Overhauser field differed depending on whether the external magnetic field was increased or decreased. This led to a noticeable difference in RSA peak positions between field upsweeps and field downsweeps. In addition, the magnitude of the Overhauser field changed with a periodicity in applied magnetic field that matched the periodicity of the electron spin polarization, but the question of how nuclear spins were polarized transverse to the optically-pumped electron spin polarization was left unanswered.

In this article, we examine these previous findings in light of a model based on the optical Stark effect, which can rotate spin polarization about the optical axis. We develop a computational model and identify its dependence on the experimental parameters of pump wavelength and magnetic field history. We demonstrate experimentally that the model describes our physical system before showing that it reproduces the magnetic field sweep results from Ref.\,\,[\onlinecite{Macmahon}]. Finally, we give an intuitive explanation for understanding this sweep direction dependence.

\section{\label{sec:model}Model}
\begin{figure*}
\includegraphics[width=15cm]{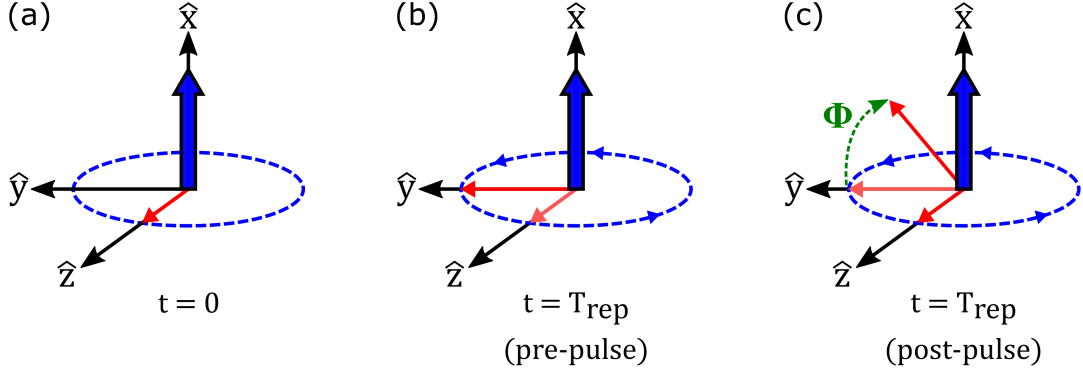}
\caption{\label{fig1}
Schematic of spin polarization generation and precession in the presence of periodic circularly polarized pump pulses. \textbf{(a)} At time $t = 0$, the first pump pulse generates an electron spin polarization (thin red arrow) along the optical axis $\hat{z}$. An external magnetic field (thick blue arrow) is applied along the transverse direction. \textbf{(b)} At time $t = T_{rep}$, just before the next pump pulse is incident on the system, the spin polarization has precessed and will lie somewhere in the plane perpendicular to the external magnetic field. \textbf{(c)} At time $t = T_{rep}$, the next pump pulse is incident on the system. The optical Stark effect manifests as a magnetic field along the optical axis experienced by the existing spin polarization. The polarization rotates about the optical axis by angle $\Phi$ (green). There is now a component of the spin polarization along the external magnetic field direction. At the same time, spin polarization is generated along the optical axis, and the total polarization is the vector sum of the two. The process outlined in this figure continues as pump pulses arrive at intervals of $T_{rep}$.
}
\end{figure*}

Electron spin polarization can be produced and studied with time-resolved optical pump-probe techniques \cite{Crooker}. Circularly-polarized light generates spin polarization along the optical axis $\hat{z}$ \cite{Meier}. We perform our experiments in the Voigt geometry, where an external magnetic field of magnitude $B_{ext}$ is applied along the $x$-axis, perpendicular to the optical axis. The spin polarization will precess about the magnetic field axis; the precession dynamics can be described classically with
\begin{subequations}
\begin{align}
S_z(t) & = S_{0}\cos\left(\Omega t\right)e^{-t/T_{2}^{*}}\\
S_y(t) & = -S_{0}\sin\left(\Omega t\right)e^{-t/T_{2}^{*}}  \\
S_x(t) & = S_{0x}e^{-t/T_{2}^{*}},
\end{align}
\end{subequations}
where the Larmor precession frequency is $\Omega = \frac{g\mu_B}{\hbar}B_{ext}$, $g$ is the electron g factor, $\mu_B$ is the Bohr magneton, $\hbar$ is the reduced Planck constant, $T_{2}^{*}$ is the electron spin dephasing time, $S_0$ is the spin polarization generated by the pump pulse along $\hat{z}$ at time t = 0, and $S_{0x}$ is the spin component along the magnetic field axis at t = 0.
It has been shown that excitation with pulsed circularly polarized light can lead to splitting of the two-fold spin degenerate excited electron state; this is known as the optical Stark effect \cite{Gupta, Economou}. This splitting can be described as an effective magnetic field along the optical axis which will rotate electron spins about this axis over the duration of the pump pulse. If the pulse duration is much less than the repetition time of the laser pulses $\left( \tau_{p} \ll T_{rep} \right)$, then the relations between the rotated electron spin components pre $\left( - \right)$ and post $\left( + \right)$ pump pulse can be described by \cite{Yugova}:

\begin{subequations}
\begin{align}
S_z^+ & = \frac{Q^2 -1}{4}+ \frac{Q^2 +1}{2}S_z^-.\\
S_y^+ & = Q\cos\left(\Phi\right)S_y^- -  Q\sin\left(\Phi\right)S_x^- \\
S_x^+ & = Q\cos\left(\Phi\right)S_x^- +  Q\sin\left(\Phi\right)S_y^- 
\end{align}
\noindent The optical Stark rotation angle $\Phi$ due to the pulse, measured in radians with respect to the $y$-axis, is given by
\begin{align}
\Phi & = \arg \left(\frac{\Gamma^2(\frac{1}{2} - i\Delta)}{\Gamma^2(\frac{1}{2} - i\Delta - \frac{\theta}{2\pi} )\Gamma^2(\frac{1}{2} - i\Delta + \frac{\theta}{2\pi} )}\right)
\end{align}
\noindent while the rotation amplitude $Q$ is given by
\begin{align}
Q & = \left|\frac{\Gamma^2(\frac{1}{2} - i\Delta)}{\Gamma^2(\frac{1}{2} - i\Delta - \frac{\theta}{2\pi} )\Gamma^2(\frac{1}{2} - i\Delta + \frac{\theta}{2\pi})}\right|.
\end{align}
\end{subequations}

Both $\Phi$ and $Q$ are dependent on the pump detuning from resonance $\Delta = \left( E_{pump} - E_{g} \right) \tau_{p} / 2 \pi \hbar$ and the pulse area $\theta = \int 2 \left| \langle d \rangle  E\!\left( t \right) \right| dt / \hbar$. A schematic of spin generation, precession, and rotation is shown in Fig. 1.

The relationship between the spin precession about the applied magnetic field and spins rotated into the field direction by the optical Stark effect can be analytically evaluated by setting $\vec{S}(t \rightarrow 0)= \vec{S}^{+}$ and $\vec{S}(T_{rep})=\vec{S}^-$ in Eqns(1) and plugging these into Eqns(2). From these relations, the spin polarization components from pumping with an infinite train of pulses separated by repetition time $T_{rep}$ can be evaluated by summing over each pulse with the remaining spin polarizations from previous pulses. This summation is a convergent geometric series whose analytic continuation can be written as

\begin{subequations}
\begin{align}
S_z(t) & = S_{0}r\cos\left(\Omega t - \gamma\right)e^{-t/T_{2}^{*}}  \\
S_y(t) & = -S_{0}r\sin\left(\Omega t - \gamma\right)e^{-t/T_{2}^{*}}  \\
S_x(t) & = -S_{0}Kr\sin\left(\Omega T_{rep} - \gamma\right)e^{-t/T_{2}^{*}},
\end{align}
\end{subequations}
where
\begin{equation*}
    K= \frac{Qe^{-T_{rep}/T_{2}^{*}}\sin\left(\Phi\right)}{1-Qe^{-T_{rep}/T_{2}^{*}}\cos\left(\Phi\right)}
\end{equation*}
\begin{equation*}
    \alpha = Q\cos\left(\Phi\right) - KQ\sin\left(\Phi\right)
\end{equation*}
\begin{equation*}
    r = \left(1 - 2\alpha e^{-T_{rep}/T_{2}^{*}}\cos\left(\Omega T_{rep}\right) + \alpha^2  e^{-2T_{rep}/T_{2}^{*}} \right)^{-1/2}
\end{equation*}
\begin{equation*}
    \gamma = -\arctan{ \left(\frac{\alpha e^{-T_{rep}/T_{2}^{*}}\sin\left(\Omega T_{rep}\right)}{1-\alpha e^{-T_{rep}/T_{2}^{*}}\cos\left(\Omega T_{rep}\right) }\right)}
\end{equation*}
\begin{equation*}
    S_{0} = \left(\frac{Q^{2} - 1}{2}\right) \frac{1}{r\cos\left(\gamma\right) - \left(\frac{Q^{2} + 1}{2}\right) r\cos\left(\Omega T_{rep} - \gamma\right) e^{\frac{-T_{rep}}{T_{2}^{*}}}}.
\end{equation*}

The spin polarization generated per pump pulse after an infinite train of pulses $S_0$ is derived by plugging Eqn(3a) for $S_z(t \rightarrow 0)= S_z^{+}$ and $S_z(t \rightarrow T_{rep})=S_z^-$ into relation Eqn(2a). When the electron spin lifetime is greater than the repetition time between pulses, the remaining spin polarization from previous pulses will constructively or destructively interfere leading to resonant spin amplification (RSA) \cite{Kikkawa1998}. Pumping on resonance with the optical transition energy $(\Delta = 0)$ will yield no optical Stark rotation $(\Phi=0)$ and Eqn(3a,b) will reduce to previously derived equations for resonant spin amplification for any time delay \cite{Trowbridge, Glazov}. These equations which describe the electron spin components of resonant spin amplification and their dependence on pump power and detuning have been previously derived \cite{Yugova2012}, however their formulation was defined for pump-probe delay $t \rightarrow 0$ and was not generalized to all time delays. Fig. 2 shows the calculated components of the spin polarization produced by periodic optical pumping and the optical Stark effect at three wavelengths. Note that $S_x$ changes sign with the sign of optical detuning.

\begin{figure*}
\includegraphics[width=17.8cm]{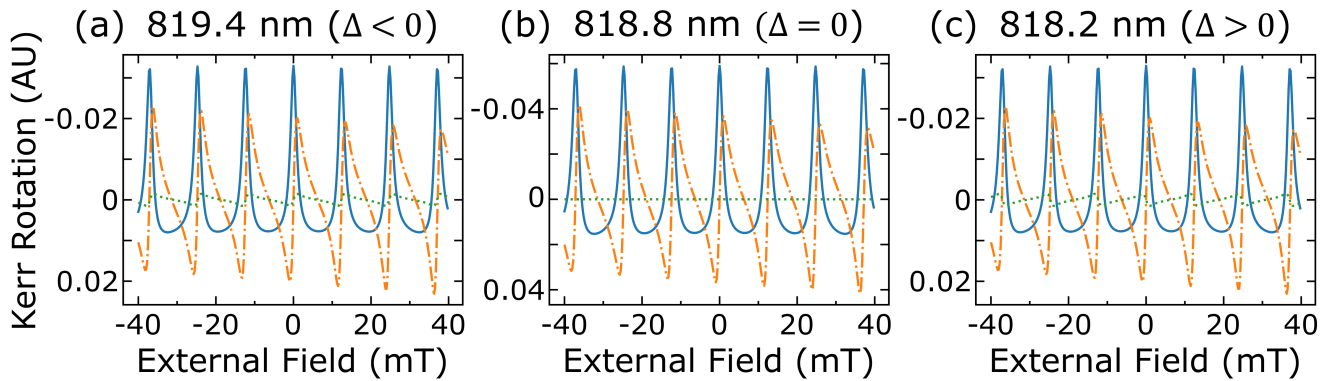}
\caption{\label{fig2}
Simulated components of electron spin polarization under periodic optical excitation and the optical Stark effect in the absence of nuclear polarization, at a time 13 ns after the arrival of the most recent pump pulse, as a function of external magnetic field, for optical wavelength (a) 819.4 nm (negative detuning), (b) 818.8 nm (resonance), and (c) 818.2 nm (positive detuning). The $z$-component of electron spin polarization $S_z$ is shown in blue (solid), while $S_y$ is shown in orange (dot-dashed), and $S_x$ is shown in green (dotted). In contrast to $S_z$ and $S_y$, $S_x$ changes sign with the sign of optical detuning and is zero on resonance. The model assumes spin lifetime $T_{2}^{*} =$ 30 ns and pulse area $\theta = \pi/4$.
}
\end{figure*}

Optically oriented localized resident carriers will couple with nuclei; this is characterized by the Fermi contact term of the hyperfine interaction. This interaction will transfer spin momentum between the electron spin and the nuclear spin system in a process called dynamic nuclear polarization. The spin momentum transfer process of nuclei for a magnetic field along $\hat{x}$  can be described by the rate equation \cite{Dyakonov, Abragam, Paget}

\begin{equation}
    \frac{dI_{av}}{dt} = - \frac{1}{T_{1e}}\left(I_{av} - \frac{4}{3}I\left(I+1\right)\langle S_x\rangle\right) - \frac{I_{av}}{T_{1n}},
\end{equation}
where $I_{av}$ and $\langle S_x\rangle$ are the time-averaged nuclear and electron spin components along the external magnetic field, respectively, $T_{1e}$ is the nuclear polarization time due to the hyperfine interaction with electrons, and $T_{1n}$ is the phenomenological nuclear relaxation time due to all other relaxation processes. For the results presented here, we set $T_{1e}$ = 180 seconds and $T_{1n}$ = 20 seconds, consistent with values reported in the literature \cite{Meier, Lu, Kaur, Koelbl}. For the experimental results presented in this paper, these long polarization and relaxation times exceed the time interval between measurements. As such, a steady-state solution to Eqn(4) does not accurately reproduce the observed behavior, and we must instead numerically solve the differential equation. 
When optically orientating electron spins with resonant excitation in the Voigt geometry, there is no distinguishable spin polarization along the magnetic field axis. However, with spectrally detuned periodic pumping, there will exist a nonzero spin polarization component rotated into the external field direction by the optical Stark effect. The time-averaged electron spin component along the magnetic field axis pumped by an infinite train of pulses can be analytically evaluated by integrating Eqn(3c) over the pulse repetition period $T_{rep}$:

\begin{equation}
    \langle S_x\rangle = \frac{T_{2}^{*}}{T_{rep}}\left( e^{-T_{rep}/T_{2}^{*}} - 1\right)KS_0r\sin\left(\Omega T_{rep}-\gamma\right).
\end{equation}

The complete time- and external magnetic field dependent behavior of the nuclear spin system polarization due to the electron spin system is described by plugging our analytic $\langle S_x\rangle$ into Eqn(4). The polarized nuclear spin system will exhibit an Overhauser magnetic field $B_n$. This additional field will add to or subtract from the external magnetic field depending on the polarization of the nuclear system ($I_{av}$). The total effective magnetic field will change the Larmor precession frequency of the electron spin system:

\begin{subequations}
\begin{align}
\Omega_{eff} & = \frac{g\mu_B}{\hbar}{B}_{tot} = \frac{g\mu_B}{\hbar}\left({B}_{ext} + B_n\right) \\
\vec{B}_{n} & = bn*I_{av}\left(t, \langle S_x\left(B_{tot}\right)\rangle\right) \hat{x}\\
bn & = \sum_i \frac{ A_i \chi_i}{\mu_B g}
\end{align}
\end{subequations}
where $A_i$ is the hyperfine coupling constant and $\chi_i$ is the abundance of particular nuclei denoted by subscript $i$. For our simulations, we use hyperfine values of $A_{\textrm{As}75}$ = 46 $\mu$eV, $A_{\textrm{Ga}71}$ = 48.5 $\mu$eV, $A_{\textrm{Ga}69}$ = 38.2 $\mu$eV \cite{Zhukov2014} and abundances of  $\chi_{\textrm{As}75}$ = 100\%, $\chi_{\textrm{Ga}71}$ = 39.9\%, $\chi_{\textrm{Ga}69}$ = 60.1\% \cite{Machlan}.
This feedback effect between the coupled electron-nuclear spin systems under periodic pumping can be fully described by the coupled equations Eqns(4,5,6). On the lab time scales of our experiments, where the time between data points is shorter than $T_{1e}$ and $T_{1n}$, the coupled spin system does not reach steady state and dynamic nuclear polarization within this regime will manifest nonlinear transient effects on the electron spin system. These nonlinear interactions can be modeled by numerically integrating the coupled equations Eqns(4,5,6). To simulate the nuclear spin polarization response from the periodic pumping of the electrons within our experiments, we develop an algorithm where we: 

\begin{enumerate}
	\item Initially set the external field ${B}_{ext}$ to the parameters of the experiment and set the initial condition nuclear polarization $I_{av}$ to zero.
	\item Numerically integrate the coupled Eqns(4,5,6) over the lab time of measurement at that external field value.
	\item Set the calculated nuclear polarization as the new initial condition for the next experimental magnetic field measurement point and numerically integrate.
	\item Repeat steps 2 and 3 until all measurement points of the experiment are replicated.
\end{enumerate}

In order to account for the dipole-dipole coupling between nuclear spins when the applied magnetic field is near zero, we also assume that the nuclear polarization is reduced following the relation \cite{Meier}

\begin{equation}
    I_{av} = I_{av,0} * B_{ext}^2/\left(B_{ext}^2 + \xi B_{L}^2\right),
\end{equation}
where $B_L$ is the local field and has been theoretically calculated for GaAs to be ~0.145 mT \cite{Paget}. Recent experiments have measured local fields on the order of 0.6 mT \cite{Litvyak}. In our modeling, we set $B_L$ = 0.1 mT and $\xi$ = 1.

Analysis of the feedback between the nuclear spin system and electrons rotated into the magnetic field axis by the optical Stark effect has been recently performed in Ref. \cite{Kopteva}. This analysis also provides a formulation of $\langle S_x(B_{tot})\rangle$ which is equivalent to the derived alternative formulation presented by Eqn(5). However, the analysis performed in Ref. \cite{Kopteva} is limited to the steady-state stable solutions of the coupled electron-nuclear system. We use our model to explore the hysteretic transient solutions which describe and qualitatively replicate the previously unexplained nonlinear magnetic field sweep dependent behavior in Ref. \cite{Macmahon}. We justify this theoretical model by performing additional optical pump-probe experiments investigating the nuclear spin polarization dependence on pump wavelength and external magnetic sweep speed.

\section{\label{sec:results}Experimental Results}

\begin{figure*}
\includegraphics[width=17.8cm]{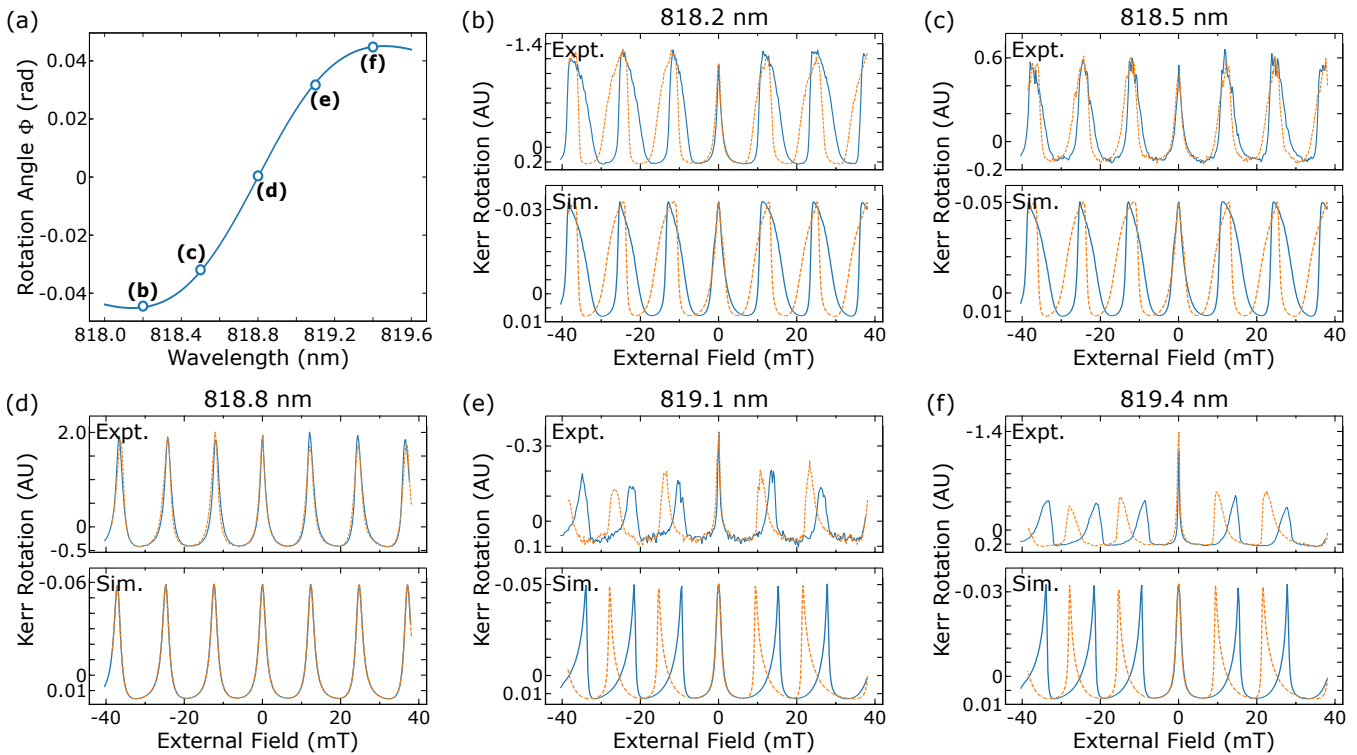}
\caption{\label{fig3}
\textbf{(a)} Rotation angle $\Phi$ of the electron spin polarization due to the optical Stark effect, as a function of the center wavelength of the incident circularly polarized laser pulse, calculated from Eqn(2d). \textbf{(b) - (f)} Above: Kerr rotation measured as a function of external magnetic field for a fixed pump-probe delay of 13 ns. Below: Simulated spin polarization magnitude as a function of external magnetic field for a fixed pump-probe delay of 13 ns. In both experiment and simulation, the field is swept from $-40$ to $+40$ mT (blue solid line) for the upsweep and from $+40$ to $-40$ mT (orange dashed line) for the downsweep. Each of (b) - (f) correspond to a different pump and probe wavelength marked in (a). Kerr rotation is directly proportional to electron spin polarization in our experiments, so the experiments can be compared qualitatively to the simulated data. 
}
\end{figure*}

As described above, the angle $\Phi$ at which spins are rotated into the $x$-axis by the optical Stark effect is dependent on the pump area and detuning. This suggests that through the coupled interactions between the spin systems described by Eqns(4,5,6), the nuclear system will acquire dependence on the laser pump wavelength. This wavelength dependence can be directly simulated by performing the modeling algorithm under the experimental parameters of external magnetic field and measurement timing. By performing this simulation for different detuning $\Delta$, we can replicate the corresponding behavior observed in our experiments.

We compare the wavelength dependence predicted by our model to our experimental results. The experimental data in this paper correspond to a 2-$\mu$m-thick Si-doped GaAs epilayer (doping density $n = 3 \times 10^{16} \ \text{cm}^{-3}$). The epilayer was grown on a 1-$\mu$m-thick undoped AlGaAs epilayer, which was grown on an undoped (001) GaAs substrate. Both epilayers were deposited by molecular-beam epitaxy. We mounted the sample in a helium flow cryostat and maintained the sample temperature at 10 K.

We optically generate electron spin polarization in the GaAs epilayer using a mode-locked Ti:S laser tuned near the bandgap of GaAs. The laser outputs a roughly 2 ps pulse every 13.16 ns (76 MHz repetition rate). We detect electron spin polarization using a pump-probe measurement scheme, recording the Kerr rotation of a reflected linearly-polarized probe pulse at a fixed time delay after the spin-generating circularly polarized pump pulse. An adjustable mechanical delay line allows us to vary this time delay and make time-dependent measurements. To facilitate this measurement, we employ a cascaded lock-in detection scheme. We modulate the pump helicity between right- and left-handed circular polarization at 50 kHz via a photoelastic modulator; we modulate the probe beam at 1370 Hz via a mechanical optical chopper. For further details on the Kerr rotation measurement technique, see Ref.\,\,[\onlinecite{Macmahon}].

We place the sample between the poles of an electromagnet, applying an external magnetic field transverse to the optical axis. For magnetic-field-dependent measurements, we fix the pump-probe time delay and vary the magnitude and direction of the external magnetic field.

For the data presented in Fig. 3, the pump-probe delay time is fixed at $-160$ ps, or 13 ns after the previous pump pulse. The initial external magnetic field is set to $-40$ mT and then increased in steps of 0.25 mT at 1.1 second intervals until the external field reaches $+40$ mT. The magnetic field is then decreased in steps of 0.25 mT at 1.1 second intervals until the field magnitude returns to $-40$ mT. Fig.  3(b-f) show the Kerr rotation signal for five laser wavelengths. The initial magnetic field upsweep ($-40$ mT to $+40$ mT) is shown by the solid blue line and the returning downsweep ($+40$ mT to $-40$ mT) is shown by the dashed orange line. Note again that the time between measurements is much shorter than the timescales of nuclear polarization and depolarization in our system.

The peaks in the data at laser wavelength 818.8 nm, shown in Fig. 3(d), resemble resonant spin amplification in the absence of dynamic nuclear polarization, and there is no discernible difference in signal between magnetic field sweep directions. This suggests that at pump wavelength 818.8 nm, there is minimal Overhauser field interacting with the electron spin system. This is consistent with the expectation that no rotation due to the optical Stark effect should occur when the laser wavelength is on resonance and, thus, no dynamic nuclear polarization occurs. Direct comparison with our model at this wavelength suggests the electron polarization lifetime $T_{2}^{*}$ is approximately 30 ns. Assuming $T_{2}^{*}$ is largely independent of pump wavelength, we can replicate the changes in peak shape produced by DNP measured at other wavelengths by adjusting the pump laser detuning in the model. Note that the effect on peak shape is different depending on the direction of laser detuning, which is consistent with the opposite sign of the Stark rotation angle $\Phi$ for positive and negative detuning. For example, in Fig. 3(f), with pump laser wavelength 819.4 nm, we observe a prolonged rising edge, relative to the data in Fig. 3(d), with the same peak warping as our previous experiments \cite{Macmahon}. However, for positive detuning, as shown in Fig. 3(b), with pump laser wavelength 818.2 nm, we observe a prolonged falling edge relative to the case with small detuning in Fig. 3(d).

Note that the experimental results measure Kerr rotation, the amplitude of which depends on laser wavelength through both spin polarization generation and detection \cite{Meier}. The numerical results plot the spin polarization $S_z$, ignoring the wavelength-dependence of detection. We observe in our measurements a difference in amplitude between the center peak near $B_{ext} = 0$ and the other peaks. This amplitude difference between the center peak and its neighbors roughly follows the detuning in sign, but it does not emerge from our model. Similarly, we failed to predict the degree to which the center peak is narrowed. This observed narrowing is most extreme for negative detuning. The width of the zero peak in magnetic field sweeps is usually indicative of the electron spin lifetime of the system, but the nuclear field can artificially narrow the center peak. Due to the dipole-dipole coupling implemented in the model via Eqn(7), the nuclear field is reduced near zero external field, though it does not fully vanish until $B_{ext} = 0$.  Thus the edges of the center peak will shift inward, but its center will stay fixed at zero external field. This is demonstrated in the simulated Kerr rotation. We suspect that these discrepancies between our observed and modeled center peak amplitudes and widths may be attributable to an electron spin lifetime that changes with wavelength, external magnetic field, or nuclear polarization.


\begin{figure*}
\includegraphics[width=17.8cm]{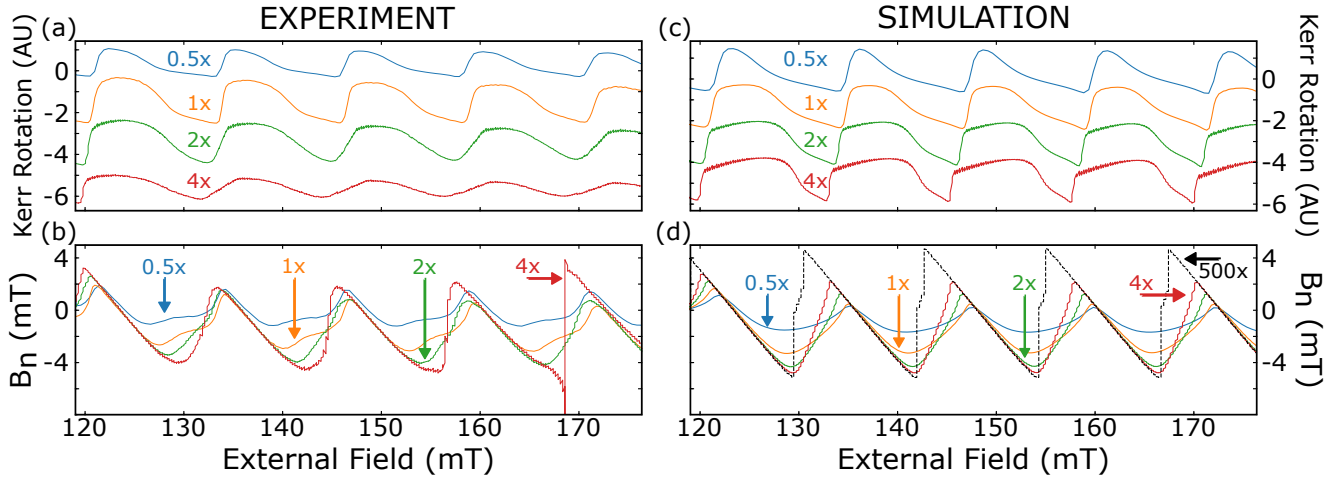}

\caption{\label{fig4}
Comparison of experimentally obtained and simulated Kerr rotation and Overhauser field for time-dependent field sweeps. \textbf{(a)} Kerr rotation measured as a function of external magnetic field for a fixed pump-probe delay of 13 ns and laser wavelength 818.2 nm. The external field is swept from 120 to 180 mT in steps of 0.25 mT. The labels indicate the number of measurements taken at each magnetic field. Here, $1\times$ corresponds to the field step timing and spacing used throughout this paper, effectively 0.23 mT/s. The exception is $0.5\times$, which utilized the same field step timing as $1\times$ but skipped every other field step. The field sweeps are offset vertically for clarity. While a pump-probe delay of 13 ns is shown, measurements were recorded for a series of delays in order to extract the Overhauser field. \textbf{(b)} Overhauser field for the external field sweep shown in (a), extracted from time- and field-resolved Kerr rotation measurements as described in the text. \textbf{(c)} Simulated spin polarization magnitude as a function of external magnetic field for a fixed pump-probe delay of 13 ns. The simulated experiment follows the same procedure and detuning outlined in (a), and the sweeps are likewise offset for clarity and comparison. \textbf{(d)} Simulated Overhauser field corresponding to the same experimental conditions. The black dashed line corresponds to an experiment with 500$\times$ the number of measurement steps at each field, effectively a steady-state measurement.
}
\end{figure*}

In performing a field sweep, the nuclear polarization’s time dependence incorporates the time spent at the measurement fields. Longer time spent at each fixed external field suggests that the nuclear system has more time to approach its steady state value. However, over the scan the nuclear polarization builds up from interactions during the previous measurement points. This build up is increased with longer measurement intervals.

To experimentally determine the nuclear system’s time dependence, we adjust the time interval between fixed field measurement points. In our previous experiments, measurements were taken for 1.1 s between incrementing field steps of 0.25 mT which is effectively a magnetic field sweep rate of 0.23 mT/s. To increase the time spent between field steps, we increase the number of measurements recorded at each external field step. For example, $2\times$ corresponds to two measurement points taken at each field value, leading to half the effective magnetic field sweep rate.

The Overhauser field can be extracted by performing field sweep scans at multiple time delays between pump and probe pulses \cite{Macmahon}. This results in two-dimensional Kerr rotation data as a function of both time delay and external field. The nuclear polarization is reset before each field scan by setting the external field to zero. This assures that magnetic field sweep dependent nuclear polarization builds to the same magnitudes at the corresponding external field values. We fit the Kerr rotation with a decaying cosine as a function of time delay $A e^{-\Delta t / T_{2}^{*}} \cos \left( \Omega_{eff} \Delta t + \gamma \right)$ separately for each field value. This provides the effective Larmor precession frequency and subtracting out the external field in accordance with Eqn(6) will give the Overhauser field. For all experimentally extracted $B_{n}$ shown, the measurements were taken at time delays between $-1920$ and 1920 ps at intervals of 160 ps, and the magnetic field was swept from 120 mT to 180 mT in steps of 0.25 mT at 1.1 s intervals. The fitting procedure treated the data measured at negative time delays as positive delay times offset by 13.16 ns. This was done to insure that there were enough oscillations in the time-resolved Kerr rotation data for fitting. The $\gamma$ term in Eqn(3) was used as the initial fitting parameter for the phase. Figures 4(a) and 4(b) shows the measured Kerr rotation at 13 ns pump-probe delay and extracted Overhauser field for effective field sweep rates of $0.5\times$, $1\times$, $2\times$, and $4\times$.

The same effective sweep rates can be implemented into the model by inputting the same array of the experimental applied field values used in Fig. 4(a). The calculated Overhauser fields shown in Fig. 4(d) qualitatively replicate the observed magnetic field sweep rate dependent peak warping behavior. Our simulations demonstrate that slower external field sweeps allow for the nuclear system polarization to build for more time between subsequent external field values. The periodic Overhauser field will alter the effective Larmor precession of the electron spin system and elongate (for positive detuning at 818.2 nm) the spin polarization peaks with respect to the changing external field. This observed sweep rate dependent distortion of the Kerr rotation data is replicated by our simulations in Fig 4(c).

\section{\label{sec:direction}Direction Dependence}
\begin{figure*}
\includegraphics[width=17.8cm]{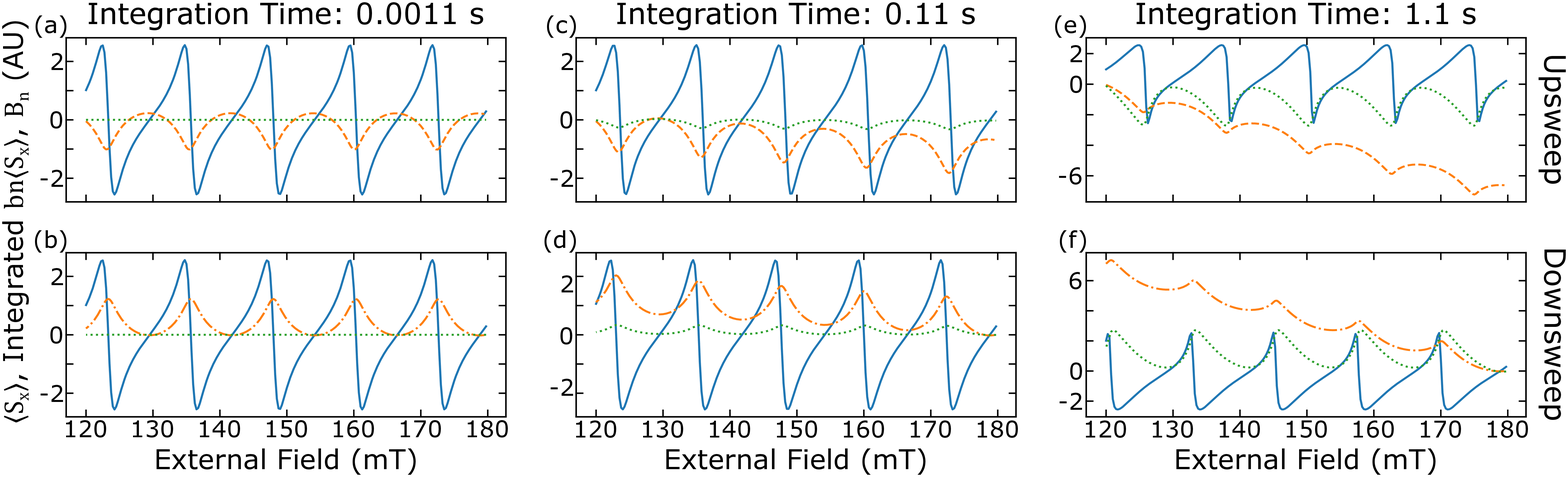}
\caption{\label{fig5}
Simulated average component of spin polarization along the external field direction $\langle S_x \rangle$ and its connection to the Overhauser field for increasing integration times. Each plot displays $\langle S_x \rangle$ as a function of external magnetic field magnitude (solid blue line), integrated $\langle S_x \rangle$ over the course of the field sweep multiplied by bn and scaled $50 \times$ (dashed orange line), and resulting Overhauser field due to $\langle S_x \rangle$ (dotted green line). At each applied magnetic field magnitude, the optical Stark effect rotates some electron spins into the direction parallel to the magnetic field. During the course of a field sweep, the change in external field magnitude changes the degree to which the spin polarization is rotated. The time duration between field steps then changes the amount of accumulated $\langle S_x \rangle$. By starting with zero net nuclear spin, the Overhauser field is proportional to this accumulation of $\langle S_x \rangle$. Since the integrated $\langle S_x \rangle$ has opposite sign for upsweeps and downsweeps, the Overhauser field takes on different character for the two directions of magnetic field sweep.
}
\end{figure*}

We have shown that the nuclear polarization changes sign dependent on whether the external magnetic field is increased or decreased \cite{Macmahon}. Analysis conducted with the OSE model can probe the nuclear system’s dependence on how the external field is changed to uncover the origin of this hysteretic behavior. Our model analytically describes the Overhauser field’s dependence on the rotated $S_x$ component of the electron spin system. This dependence can be seen by rewriting the nuclear rate equation Eqn(4) in terms of $\langle S_x \rangle$, $\frac{dI_{av}}{dt} + \left(\frac{1}{T_{1e}}+ \frac{1}{T_{1n}}\right) I_{av} = \frac{1}{T_{1e}} \frac{4}{3}I\left(I+1\right)\langle S_x\rangle. $
 We see that when $I_{av}=0$, the growth of the nuclear polarization follows the time averaged $S_x$ value at that particular external field. Integrating Eqn(4) as a function of changing field suggests that, over the duration of the field sweep, the nuclear polarization will cumulatively build and will increase or decrease depending on the value of the integrated $\langle S_x \rangle$. We can modify the model to allow for negligible time for nuclear polarization to build and show in Fig. 5(a) and 5(b) that the sign of the integrated $\langle S_x \rangle$ component depends on whether the external field is increasing or decreasing $(dB_{down} = -dB_{up})$.
We can further investigate the nuclear polarization's coupled dependence on integrated $\langle S_x \rangle$ and time by modifying both the integration time of our model and the magnetic field sweep direction. Numerical integration intervals of 0.11 s (Fig. 5(c) and 5(d)) and our experimental time interval 1.1 s (Fig. 5(e) and Fig. 5(f)) are both much less than the nuclear $T_{1e}$ time and allow more time for the nuclear polarization to increase in magnitude. Comparing upsweeps (Fig. 5(a,c,e)) to downsweeps (Fig. 5(b,d,f)), we see that the sign of that increase depends on the sign of the integrated $\langle S_x \rangle$.

\section{\label{sec:conclusion}Conclusion}

In summary, we show that the optical Stark effect model can explain the previously unexplained hysteretic field sweep direction dependent behavior of the nuclear spin system in GaAs \cite{Macmahon}. Dynamic nuclear polarization due to spins excited by a periodic train of pulses and rotated by the optical Stark effect has been previously observed and characterized in ZnSe \cite{Zhukov}. In those studies, the nuclear system’s $T_1$ time was orders of magnitude shorter than the lab time of their measurements \cite{Greilich, Heisterkamp2016}. Hence, their nuclear system quickly reached steady state during their experiments and no field sweep direction dependence was observed. In our system, the long $T_{1e}$ of our system allows for the nuclear polarization to continue building over the duration of the field scans. The polarization increases or decreases depending on the integrated $S_x$ component with respect to the change of field. This model lends insight on the dynamics of the coupled electron-nuclear spin system and how it can be manipulated with optical pulses. 

\begin{acknowledgments}
J.R.I. was supported by the Department of Defense through the National Defense Science and Engineering Graduate Fellowship (NDSEG) program. The work at the University of Michigan is supported by the National Science Foundation under Grant No. DMR-1607779. Sample fabrication was performed at the University of Michigan Lurie Nanofabrication Facility.
\end{acknowledgments}

\end{document}